\documentclass[twocolumn,showpacs]{revtex4}
\usepackage{epsfig}
\usepackage{amssymb,amssymb}
\usepackage{cases}
\usepackage{graphicx}
\usepackage{amssymb}
\usepackage{epstopdf}
\begin{document}
\draft

\title{Polymer chains in confined spaces and flow-injection problems: some remarks}
\author{T. Sakaue, E. Rapha\"el}
\affiliation{Laboratoire de Physique de la Mati\`ere Condens\'ee, FRE 2844 du CNRS, Coll\`ege de France, 
11 Place Marcelin Berthelot, 75231 Paris Cedex 05, France}

\begin{abstract}

We revisit the classical problem of the behavior of an isolated linear polymer chain in confined spaces, 
introducing the distinction between two different confinement regimes (the {\it weak} and the {\it strong} confinement regimes, respectively).
We then discuss some recent experimental findings concerning the partitioning of individual polymers into protein pores. 
We also generalize our study to the case of branched polymers, and study
the flow-injection properties of such objects into nanoscopic pores,
for which the  strong confinement regime plays an important role.
\end{abstract}


\maketitle

\section{Introduction}
In recent years, much attention has been paid to the structure and dynamics 
of polymer chains in confined spaces \cite{deGennes,Vilgis_2,nanopore_exp3}. 
For instance, understanding how biopolymers migrate through a narrow passageway 
to get to their targeted destination \cite{Kasiano,LubNelson,Mutu,MellerNivon,Chuang} is crucially important 
in cell biology \cite{Alberts}.
In confined spaces, the number of available configurations for a polymer chain is reduced,
leading to a free energy excess \cite{Daoud_deGennes}.
For an ideal linear chain confined in a space of characteristic size $D$ (with 
$D < R_0$, where $R_0 = aN^{1/2}$ is the natural size of the chain, $N$ is the number of monomers and $a$ is the monomer size),
this free energy excess was calculated by Cassasa \cite{Cassasa}:
\begin{eqnarray}
\frac{F_{{\rm{id}}}}{k_B T}  \; \simeq  \; \left(\frac{R_0}{D}\right)^2
\label{F_ideal}
\end{eqnarray}
(where $k_B T$ is the thermal energy).

Note that the scaling relation (\ref{F_ideal}) holds equally for an ideal chain confined between 
two walls, in a cylindrical pore or in a spherical cavity.
For a linear chain with excluded volume interactions, however, 
the confinement into a spherical cavity leads to 
an excess free energy that differs from the one corresponding
to the confinement between two walls or in a cylindrical pore.
Indeed, the nature of the confinement depends on whether the interactions between monomers become substantial or not in the confined state.

In the present article, we revisit these issues for various molecular
architectures and propose to discriminate between two
confinement regimes (depending on the geometrical nature of the confinement).
In the first regime, called the {\it weak confinement regime} (WCR), the configuration of the polymer is restricted due to the confinement, but the conformation of the polymer at large length scale is still somewhat dilute.
The free energy required for the confinement (measured from the unperturbed state in bulk) is of purely entropic origin, and is an extensive function of the monomer number $N$.
In the second regime, called the {\it strong confinement regime} (SCR), the chain is highly compressed and the confinement becomes stronger and stronger as the chain length increases.
Here, the segmental interactions dominate the entropic cost due to the confinement and, consequently, the free energy is no longer simply proportional to $N$. This departure of the excess free energy from a linear dependence on $N$  was first noticed by Grosberg and Khokhlov \cite{RedBook} in the case of a linear polymer inside a closed cavity; unfortunately, this point is sometimes overlooked in the current polymer literature.
We will show that for a linear chain, the SCR is only obtained in the case of a spherical cavity, while confinement between 
two walls or in  a cylindrical pore corresponds to the WCR (Sections II.A and II.B).
We then discuss some recent experimental findings concerning the partitioning of 
poly(ethylene glycol) (PEG) molecules inside protein pore is \cite{aHL,aHL2,aHL3,aHL4,aHL5,nanopore_exp3} (Section II.C).
We next generalize our study to the case of polymers with arbitrary connectivity, using the concept of 
spectral dimension (Section III). We show in particular that for a branched polymer, the SCR
is already realized in a cylindrical pore. This is closely related to the notion of a "minimum diameter" for
the capillary 
(introduced by Vilgis {\it et al.} \cite{Vilgis_1,Vilgis_2}) and should be properly taken into account when one 
studies flow-injection problems \cite{Gay,deGennes_review} (Section IV).

\section{Linear polymer in confined spaces}
\subsection{Linear chain in the WCR regime}
Although in many practical cases the confinement geometry may be quite complicated, 
many fundamental features can be understood by considering simple geometries such as slits, capillaries and spherical cavities.
If a polymer chain is brought into such a confined space, the number of available spatial dimensions (which is $d = 3$ in bulk) is reduced at large scales.
Hereafter, we will denote by $d_c$ the number of "subtracted dimensions": at large length scales, the chain behaves as
in a space of $d=3 - d_c$ dimensions. For example, for a chain confined into a slit, one has $d_c = 1$.
\begin{figure}[ht]
\begin{center}
\includegraphics[width=8cm]{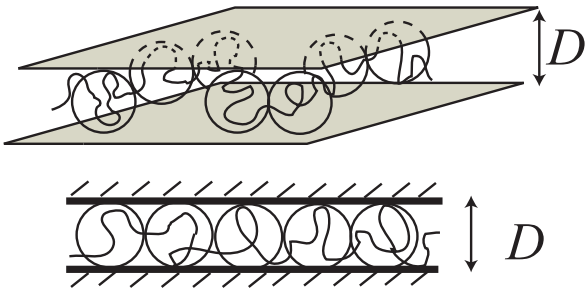}
\caption{Drawing of a linear chain confined between a slit of width $D$ ({\it top}), and inside a cylindrical pore
of diameter $D$ ({\it bottom)} (after \cite{deGennes}).}
\label{linear_slit_tube}
\end{center}
\end{figure}
For the sake of simplicity, let us first consider the case of linear polymer. 
In dilute solution, its radius is given by \cite{deGennes}
\begin{eqnarray}
R_{3} = a N^{\nu_3}
\end{eqnarray}
where $\nu_3$ is the Flory exponent (in a good solvent, $\nu_3 \simeq 3/5$).
If the chain is brought into a slit of diameter $D$ ($d_c = 1$), its conformational behaviours depend on the ratio $R_{3}/D$.
If $D < R_{3}$, the effect of confinement is important and the chain displays a two-dimensional behavior at length scales 
larger than $D$ (Fig. \ref{linear_slit_tube}: top) \cite{Daoud_deGennes,deGennes}.
The polymer can be envisioned as a sequence of blobs of size $D$ in the plane.
The overall shape of the polymer is thus characterized by a thickness $D$ and a radius $R_{2}$
given by  \cite{deGennes}

\begin{eqnarray}
R_{2} \simeq D \left( \frac{N}{g}\right)^{\nu_2}
\label{R_2d}
\end{eqnarray}
with $\nu_2$ being the Flory exponent in two-dimension ($\nu_2 \simeq 3/4$ in a good solvent condition).
The number of segments in a blob $g$ is given by
\begin{eqnarray}
g \simeq \left(\frac{D}{a}\right)^{\frac{1}{\nu_3}}
\label{segment_in_blob}
\end{eqnarray}
Equation (\ref{segment_in_blob}) means that inside blobs, the effect of wall is insignificant, thus, the internal correlations between segments are identical to those in bulk.
From Eqs. (\ref{R_2d}) and (\ref{segment_in_blob}), the radius of a linear chain confined in a slit 
is thus given by
\begin{eqnarray}
R_{2} \simeq a N^{\nu_2} \left(\frac{a}{D}\right)^{{\frac{\nu_2}{\nu_3}} - 1}
\end{eqnarray}
In good solvent conditions, the preceding equation yields
\begin{eqnarray}
R_{2} \simeq  a  N^{\frac{3}{4}} \left(\frac{a}{D} \right)^{\frac{1}{4}}  
\end{eqnarray}

Now, let us consider the free energy per chain measured from the unperturbed state in bulk solution.
As described above, the chain has to be reflected at the wall.
For each collision, there exists a entropy cost of the order of thermal energy $k_BT$.
Therefore, the free energy required for the chain confinement is easily evaluated as the entropy loss due 
to the confinement of $N/g$ blobs \cite{deGennes}:
\begin{eqnarray}
\frac{F_2}{k_BT} \simeq N \left( \frac{a}{D} \right) ^{\frac{1}{\nu_3}}
\label{F_slit}
\end{eqnarray}
In good solvent conditions, on gets
\begin{eqnarray}
\frac{F_2}{k_BT} \simeq N \left( \frac{a}{D} \right) ^{\frac{5}{3}}
\label{F_slitgood}
\end{eqnarray}
Note that the above confinement free energy can also be deduced from the following scaling argument \cite{Daoud_deGennes,deGennes}:
(a) $F_2/(k_BT)$ is dimensionless and depends only on the length ratio $R_{3}/D$, (b) the leading term in $F_2/(k_BT)$ 
should be an extensive function of $N$. By imposing these two conditions, one indeed recovers Eq. (\ref{F_slit}).

A similar argument can be applied to the chain in a thin capillary (of diameter $D < R_{3}$), where the polymer shows one-dimensional 
($d_c = 2$) behaviours at large length scale (Fig. \ref{linear_slit_tube}: bottom).
Here, the chain can be envisioned as a succession of blobs of size $D$.
At length scales larger than $D$, the chain shows one-dimensional conformational statistics, which results in
\begin{eqnarray}
R_{1} \simeq a N \left(\frac{a}{D}\right)^{{\frac{1}{\nu_3}} - 1}
\end{eqnarray}
Thus, in good solvent conditions, we find
\begin{eqnarray}
R_{1} \simeq  a N \left(\frac{a}{D}\right)^{\frac{2}{3}}
\end{eqnarray}
The free energy $F_1$ required to confine the chain inside the capillary is here again of order $k_B T$ per blob, leading to
\begin{eqnarray}
F_1 \simeq F_2
\label{F12}
\end{eqnarray}
(omitting numerical prefactors of order one). In other words, the confinement free energy for a chain
in a capillary obeys the same scaling law that for a chain in slit.

\subsection{Linear chain in the SCR regime}
\label{linear_strong}
\begin{figure}[ht]
\begin{center}
\includegraphics[width=8cm]{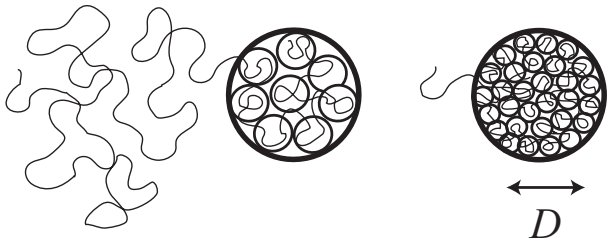}
\caption{Drawing of a linear chain progressively confined inside a spherical cavity. As the confinement takes place,
the monomer volume fraction inside the cavity increases and, consequently, 
the blob size decreases (after \cite{RedBook}).}
\label{linear_sphere}
\end{center}
\end{figure}

Let us further decrease the number of available spatial dimensions by considering the behavior of a 
linear chain confined inside a spherical cavity ($d_c = 3$).
Although sometimes adopted in recent publications, the assumption that the free energy
has still the scaling form displayed by Eq. (\ref{F_slit})  is incorrect.
This can be seen rather easily 
if one imagines the consequence of doubling the chain length (Fig. \ref{linear_sphere}).
For a chain in a slit or in a capillary, this does not affect the local properties at all, but just adds the same copy as the original system. 
Therefore, the free energy is proportional to $N$.
However, for a chain in a spherical cavity, doubling the chain length inevitably increases the volume fraction of the chain.
Therefore, the $N$ dependence of the free energy should be stronger than linear \cite{RedBook}.

The scaling form for the free energy of a chain in a closed cavity $F_0$ of size $D$ is deduced from the following thermodynamic requirements \cite{RedBook} :
(a) $F_0/(k_BT)$ is dimensionless and depends only on the length ratio $R_{3}/D$, (b) $F_0/(k_BT)$ is extensive, that is to say, under the variable transformation $N \rightarrow k N$ and $\Omega (\simeq D^3) \rightarrow k \Omega$ (with $k>0$), the free energy is modified as 
$F_0 \rightarrow k F_0$, (or equivalently, the pressure $P = -\partial F/\partial V$ remains unchanged).
From these requirements, we obtain
\begin{eqnarray}
\frac{F_0}{k_BT} \simeq \left( \frac{R_{3}}{D}\right)^{\frac{3}{3 \nu_3 - 1}}
\label{F_sphere}
\end{eqnarray}
For a polymer chain in good solvent ($\nu_3 \simeq 3/5$)
\begin{eqnarray}
\frac{F_0}{k_BT} \simeq N^{\frac{9}{4}} \left( \frac{a}{D}\right)^{\frac{15}{4}} 
\label{F_sphere_good}
\end{eqnarray}

Eq. (\ref{F_sphere}) can be rewritten by using the volume fraction of the chain $\phi \simeq Na^3/D^3$.
\begin{eqnarray}
\frac{F_0}{k_BT} \simeq N \phi^{\frac{1}{3 \nu_3 - 1}}
\label{F_sphere_phi}
\end{eqnarray}
The free energy per segment depends only on the volume fraction, which indicates the analogy with semidilute polymer solutions.
Along this line, it is possible to construct a blob picture for the chain confined in a spherical cavity.
At short distances (smaller than the blob size $\xi$), the monomer-monomer correlations are similar to those
in the bulk. The number $g$ of monomers inside a blob is thus given by $\xi \simeq a g^{\nu_3}$.
Moreover, the confined chain can be viewed as a compact stacking of blobs : $\phi \simeq g a^3/\xi^3$.
From these two relations, the correlation length is found to be \cite{RedBook}
\begin{eqnarray}
\xi \simeq a \left( \frac{D^{3}}{a^3 N}\right)^{\frac{\nu_3}{3 \nu_3 - 1}}
\label{correlation_length}
\end{eqnarray}
In good solvent conditions, the correlation length thus decays as $N^{-3/4}$.
The free energy can be estimated by using the $k_B T$ per blob ansatz:
\begin{eqnarray}
\frac{F_0}{k_BT} \simeq \frac{D^3}{\xi^3}
\label{F_sphere2}
\end{eqnarray}
Substituting Eq.(\ref{correlation_length}) into Eq.(\ref{F_sphere2}), one indeed recovers Eq.(\ref{F_sphere}).

We see that there are two qualitatively different confinement regimes, depending on the confinement geometry.
The threshold is obtained by calculating the volume fraction of the chain as a function of $N$
\begin{eqnarray}
\phi = \frac{Na^3}{(R_{d})^d D^{d_c}}\sim N^{\alpha}
\label{volume_fraction_linear}
\end{eqnarray}
where the exponent $\alpha$ is given by
\begin{eqnarray}
\alpha = \frac{2 (d_c-2)}{5-d_c}
\end{eqnarray}

If $\alpha$ is negative, as in the case of a chain confined into a slit (for which $\alpha = -1/2$), the confined chain is 
a rather dispersed system and its density decreases with chain length.
We call such a regime the {\it weak confinement regime} (WCR).
On the other hand, in the {\it strong confinement regime} (SCR), $\alpha$ is positive (as in the chain in a spherical cavity
for which $\alpha = 1$); 
the confined chain is then reminiscent of a semidilute polymer solutions and its density increases with chain length.
We see that a linear chain confined inside a cylindrical pore corresponds to the critical situation $\alpha =0$;
the overall chain is a dense packing of blobs, but its free energy is still given by that of the WCR (corresponding to entropy reduction).

Let us end this section with several remarks: 

\indent (a) If a polymer is confined into the sphere of size 
\begin{eqnarray}
D^{min} \simeq a N^{\frac{1}{3}}, 
\label{minimum_D_linear}
\end{eqnarray}
the volume fraction becomes unity.
It means that it is physically impossible to confine the polymer into the sphere smaller than the minimum size $D^{min}$.
The presence of minimum allowable size, which is much larger than the monomer size $a$, is a unique characteristic of the SCR.

\indent (b) In case of the repulsive wall (no polymer adsorption), the monomer concentration should be reduced in the vicinity of the wall and become zero at the wall.
The range of this monomer depletion is on the order of $\xi$.
This leads to a surface energy
\begin{eqnarray}
F_{{\rm surf}} \simeq k_BT D^2/\xi^2 \simeq (D/a)^{2/(1 - 3 \nu_3)}N^{2\nu_3/(3 \nu_3 - 1)}
\label{surf}
\end{eqnarray}
which is associated with the conformational entropy of the chain \cite{Lifshitz}.
In good and $\theta$ solvent conditions, $F_{{\rm surf}} \simeq k_BT (a/D)^{5/2} N^{3/2}$ and $F_{{\rm surf}} \simeq k_BT (a/D)^{4} N^{2}$, respectively.
The confinement free energy (Eq. (\ref{F_sphere_good})) arising from segmental interactions is always larger than the surface energy.

\indent (c) Self-consistent field theory results \cite{Muth3} show
that the energy of a polymer chain confined in a sphere of radius $D$ scales (for good solvent conditions) as $\sim a^3 N^2/D^3$.
Note that this last result - which is essentially a mean-field result - differs from our scaling prediction Eq.(\ref{F_sphere_good}). 

\indent (d) The behaviours of confined polymer chains under $\theta$-solvent conditions (Flory exponent $\nu =1/2$) 
differs from those of ideal chains \cite{Pincus}.
This is most prominent for strongly confined polymers.
The free energy of the confined chain in a spherical cavity is given by
\begin{eqnarray}
 \frac{F_0^{\theta}}{k_BT} \simeq \left( \frac{a}{D}\right)^{6} N^{3}
\label{F_sphere_theta}
\end{eqnarray}
The $N$ dependence is stronger than that in good solvent (Eq. (\ref{F_sphere_good})).
This is due to the fact that in $\theta$ solvent, the repulsive energy comes from the three body interactions.

\indent (e) Since the static behavior of a linear chain confined in a spherical cavity is reminiscent of a 
semi-dilute solution, we might expect that this analogy would also concerns some dynamical properties
of the chain.

\subsection{Partition coefficient and comparison with experiment}
The confinement free energy $F_c$ (measured from the unperturbed state in a the bulk) is accessible experimentally by measuring
 the equilibrium partitioning of polymer molecules between the confined space and the bulk solution.
The partition coefficient, $p$, is defined as the ratio of concentration between these two regions.
The confinement free energy is then related to the partition coefficient through
\begin{eqnarray}
p \simeq \exp{\left(-\frac{F_c}{k_BT}\right)}
\label{p_coeff}
\end{eqnarray}
For a linear chain in presence of a slit, or of a cylindrical pore, the partition coefficient is obtained using the WCR confinement 
free energy (Eq. (\ref{F_slit})): 
\begin{eqnarray}
p \simeq \exp{\left[ -\left( \frac{R_{3}}{D}\right)^{\frac{1}{\nu_3}}\right]} = \exp{\left[-\left( \frac{a}{D}\right)^{\frac{1}{\nu_3}} N\right]}
\label{p_weak}
\end{eqnarray}
In the case of a spherical cavity, on has to use the SCR confinement free energy (Eq. (\ref{F_sphere})), leading
to a partition coefficient:
\begin{eqnarray}
p \simeq \exp{\left[ -\left( \frac{R_{3}}{D}\right)^{\frac{3}{3 \nu_3 - 1}}\right]} = \exp{\left[-\left( \frac{a}{D}\right)^{\frac{3}{3 \nu_3 - 1}} N^{\frac{3 \nu_3}{3 \nu_3 - 1}} \right]}
\label{p_strong}
\end{eqnarray}
In references \cite{aHL,aHL2,aHL3}, the partitioning behaviours of poly(ethylene glycol) (PEG) molecules was investigated using single nanometer-scale pores formed by protein ion channels (see also \cite{nanopore_exp3}).
By monitoring the ionic conductance of a channel for various chain lengths, one can determine how the partition coefficient $p$ 
varies as a function
of $N$;  using Eq.(\ref{p_coeff}), one then obtains the behavior of the confinement as a function of chain length, $F_c (N)$.
From their experimental data, the authors of \cite{aHL,aHL2,aHL3} conclude that the $N$-dependance 
of the free energy is sharper than predicted by theory.
More precisely, the scaling theory for a linear chain confined in a cylindrical pore
gives $F_c \sim N$ (see Eq.(\ref{p_weak})), while the experimental data are well fitted
by a relation of the form $F_c \sim N^{\mu}$ with an exponent $\mu = 3.1 \pm 0.2$.

\begin{figure}[ht]
\begin{center}
\includegraphics[width=5cm]{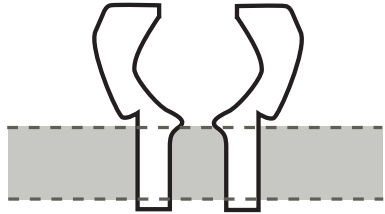}
\caption{Schematic cross section of {\it Staphylococcus aureus} $\alpha$-hemolysin channel. For more detailed information based on X-ray studies,
see \cite{nanopore_exp3,nanopore_exp2}.}
\label{protein_channel}
\end{center}
\end{figure}

Let us remark, however, that the protein pore used in these experiments 
({\it Staphylococcus aureus} $\alpha$-hemolysin ($\alpha$-toxin)) has following geometric characteristics (Fig. \ref{protein_channel}) \cite{nanopore_exp3,nanopore_exp2}.
 The channel consists of a stem region and a large cap domain.
In the cap domain, there is a pore entrance ($\sim$ approximately $2.6$ nm in diameter).
After that entrance, the diameter of the pore cross section gradually increases.
After the widest part ($\sim$ approximately $3.6$ nm in diameter), the diameter again decreases and reaches the narrowest constriction ($\sim$ approximately $1.5$ nm in diameter), which is followed by the stem (approximated by cylindrical shape with average diameter of $2.2$ nm).

Hence, if the polymer of size comparable to (or slightly larger than) the pore size enters the pore from the {\it cis} side (cap domain), it would be quite improbable for the polymer to negotiate the narrowest constriction, which effectively works as a geometric barrier.
Taking account these features into account, it may be more favorable to model the protein pore as a (almost) closed cavity.

For a linear polymer confined inside a spherical cavity, we have seen that the free energy is given by Eq. (\ref{F_sphere}).
For good solvent conditions ($\nu_3 \simeq 3/5$), we thus 
get $\mu = 9/4$ (see Eq. (\ref{F_sphere_good})), which is indeed much larger than the value $\mu = 1$ expected for 
a chain inside a cylindrical pore.
The fact that the value of $\mu$ found experimentally ($\mu = 3.1 \pm 0.2$) is larger than $9/4$
might be do to finite size effects. Indeed, if the chains are not long enough 
(and if the solvent is good, but not athermal), their radius 
of gyration might be characterized by an effective exponent $\nu$ somewhat smaller the $3/5$,
leading to a larger value of $\mu$ larger than $9/4$.

One should note, however, that other experiments suggest that the partitioning of PEG into a protein pore may be described by Eq. (\ref{p_weak}) \cite{aHL4,aHL5}.
But as mentionned by the authors, the data analyses for the cap domain is less convincing than those for the stem region, which is approximated by cylindrical shape.

We also note the other recent experiment of relevance, where behaviours of fluorescently labelled single DNA molecules are observed within spherical cavities prepared by the colloidal templating method \cite{DNA_partitioning_model_pore}.
The results indicate the importance of segmental interactions, which is in accordance with our argument for the SCR.

In summary, the partitioning behaviours of flexible polymers into a nanoscale protein pore is not clear yet.
Many experiments suggest that it may be important to take real geometry of the pore into account, and we hope that our remark may shed some light on the problem under dispute.

\section{Confined polymers with arbitrary connectivity}
In the previous section, we restricted our attention to the linear polymers only.
However, the behaviours of polymers with higher connectivity {\it i.e.}, branched polymers, in confined space are also important.
In this section, we extend the previous results to the case of polymeric fractal objects with arbitrary connectivity in a good solvent condition.

\subsection{Criterion for the confinement regime}
A simple and useful approach to confined branched polymers is a Flory type of calculation, which allows us to obtain the spatial size of branched polymers in a given geometry \cite{Vilgis_1,Vilgis_2}.
First, let us define the notation in this section; $N$ is the number of monomers in an arbitrary direction, $M$ is the total number of monomers in a given polymer. These two quantities are related to so called spectral dimension $d_s$ \cite{Alexander}: $M = N^{d_s}$. 
The physical meaning of $d_s$ would be recognized by checking some simple cases as followings: $d_s = 1$ corresponds to the simplest example of linear polymers, $d_s=2$ stands for a polymeric sheet. In more general cases, $d_s$ is non-integer and statistically branched polymers correspond to $d_s \simeq 4/3$.
Note that the spectral dimension depends on the connectivity (chemical structure of the cluster), but does not depend on the spatial conformation of the cluster.
As for the previous section, $d_c$ denotes the confining dimension (the number of subtracted dimension due to the confinement), $D$ denotes the characteristic spatial size of the confinement geometry and $R_{d}$ denotes the characteristic size of the polymer in $d$ dimensional space: the usual Flory radius for $d=3$, the radius of pancake for $d=2$ ($d_c =1$), the length of cylinder for $d=1$ ($d_c=2$).
Along with ref. \cite{Vilgis_1, Vilgis_2}, we assume $d_s<2$ in the following discussions.

Starting from the generalized Edwards Hamiltonian for the cluster with arbitrary connectivity, the corresponding mean-field Flory free energy is expressed as \cite{Vilgis_1,Vilgis_2}
\begin{eqnarray}
\frac{F}{k_BT} \simeq \frac{(R_{3})^2}{a^2 N^{2-d_s}} + \frac{a^3 N^{2d_s}}{(R_{3})^{3}}
\end{eqnarray}
After minimization of the above equation with respect to $R_{3}$, one obtains the correct fractal dimension for polymers with arbitrary spectral dimension.
\begin{eqnarray}
R_{3} \simeq a N^{\frac{d_s+2}{5}} = a M^{\frac{d_s+2}{5d_s}}
\label{branch_R}
\end{eqnarray}

If the branched cluster is confined in $d = 3 - d_c$ dimensional space, the corresponding mean-filed Flory free energy is modified
\begin{eqnarray}
\frac{F}{k_BT} \simeq \frac{(R_{d})^2}{a^2 N^{2-d_s}} + \frac{a^3 N^{2d_s}}{D^{d_c}(R_{d})^{3-d_c}}
\end{eqnarray}
If the confining dimension $d_c$ is smaller than the space dimension, the branched cluster finds a most preferable conformation in the restricted space.
After minimization with respect to $R_{d}$, one obtains the optimum size \cite{Vilgis_1,Vilgis_2}
\begin{eqnarray}
R_{d} \simeq \left( \frac{a^5 N^{d_s+2}}{D^{d_c}}\right)^{\frac{1}{5-d_c}}
\label{Branch_R_confine}
\end{eqnarray}
This leads to the volume fraction
\begin{eqnarray}
\phi = \frac{N^{d_s} a^3}{(R_{d})^{3-d_c} D^{d_c}}\sim M^{\alpha}
\label{volume_fraction}
\end{eqnarray}
where the exponent is
\begin{eqnarray}
\alpha = \frac{2(d_s-3+d_c)}{(5-d_c)d_s}
\label{exponent_beta}
\end{eqnarray}
Therefore, the critical confinement dimension is given by
\begin{eqnarray}
d_c^* =3-d_s
\end{eqnarray}
As we have already seen, for a linear polymer ($d_s=1$), the critical geometry corresponds to the capillary ($d_c=2$).
However, for branched polymers ($d_s >1$), we encounter the SCR regime already in the capillary.
Vilgis pointed out the presence of the minimum tube diameter $D^{min}$ for branched polymers, which is, in fact, closely related to the SCR \cite{Vilgis_1,Vilgis_2}.
The volume fraction of the branched polymer confined in the capillary of size $D^{min}$ becomes unity, thus, the capillary with the size $D<D^{min}$ is inaccessible to branched polymers.
From Eq. (\ref{Branch_R_confine}) and (\ref{volume_fraction}), it is found that
\begin{eqnarray}
D^{min} = a M^{\frac{d_s-1}{2d_s}}
\end{eqnarray}
For a linear polymer ($d_s = 1$), the minimum size of the capillary is equal to the monomer size as expected.
However, for branched polymers with $d_s>1$, the minimum size is much larger than the monomer size, and we notice once more that this is a unique characteristic of the SCR.
Confined in the capillary with minimum diameter, the branched polymer assumes its maximally stretched state $R_1^{max} = aN = a M^{1/d_s}$, which is much smaller than the total length of the chemical path $aM$.


\subsection{Confinement free energy of polymeric fractals}
\label{confine_branch}
In this subsection, we give the confinement free energy of fractal objects with arbitrary connectivity beyond the mean-field approximation.
The partition coefficient between a bulk and a confined space with a given geometry is obtained using Eq. (\ref{p_coeff}).

{\it Confinement in a slit}:
From the previous subsection, fractal objects with ($1 \le d_s <2$) is weakly confined in a slit geometry.
Therefore, the correlation length is set by the separation between slit $D$, and utilizing the usual scaling argument that the free energy should be proportional to the total polymerization index $M$, one obtains 
\begin{eqnarray}
\frac{F_2}{k_BT} \simeq \left(\frac{R_{3}}{D}\right)^2 \simeq \left( \frac{a}{D}\right)^2 M
\label{statistically_branch_F_slit}
\end{eqnarray}

{\it Confinement in a capillary}:
In a narrow capillary, however, branched polymers with ($1 \le d_s$) confined strongly.
To calculate the corresponding free energy, it is convenient to apply the semidilute solution analogy, as described in Sec. \ref{linear_strong} in the case of linear polymers.
A branched polymer inside the capillary is conceived as a dense piling of blobs of size $\xi$, which indicates the following relation;
\begin{eqnarray}
\frac{g}{\xi^3} \simeq \frac{M}{D^2 R_{1}}
\label{statistically_branch_blob_capillary}
\end{eqnarray}
with $g$ being the number of monomers inside a blob.
The streached length along the tube axis $R_{1}$ is deduced from Eq. (\ref{Branch_R_confine});
\begin{eqnarray}
R_{1} \simeq \left( \frac{a^5 M^{\frac{d_s+2}{d_s}}}{D^2} \right)^{1/3}
\label{statistically_branch_R_F_1_1}
\end{eqnarray}
Inside a blob, the confinement is a weak perturbation, thus, the blob size $\xi$ is related to the number of monomers $g$ inside blob through Eq. (\ref{branch_R});
\begin{eqnarray}
\xi \simeq a g^{\frac{d_s+2}{5d_s}}
\label{sratistically_branch_xi_g_1}
\end{eqnarray}
From Eq. (\ref{statistically_branch_blob_capillary}), (\ref{statistically_branch_R_F_1_1}) and (\ref{sratistically_branch_xi_g_1}), the blob size is deduced
\begin{eqnarray}
\xi &\simeq& \left( \frac{D^{2(d_s+2)}}{R_{3}^{5(d_s-1)}} \right)^{\frac{1}{3(3-d_s)}} \\
&\simeq& \left[ \frac{D^{2(d_s+2)}}{a^{5(d_s-1)}M^{\frac{(d_s-1)(d_s+2)}{d_s}}}\right]^{\frac{1}{3(3-d_s)}}
\label{sratistically_branch_xi_capillary_1}
\end{eqnarray}
For linear polymers ($d_s=1$), one recovers the conventional result $\xi \simeq D$, {\it i.e.}, blob size is set by the capillary size).
However, for polymers with higher connectivity ($d_s >1$), the correlation length decays with the increase in $M$.
The confinement free energy is evaluated by assigning $\sim k_BT$ per blob;
\begin{eqnarray}
\frac{F_1}{k_BT} \simeq \frac{D^2 R_{1}}{\xi^3} \simeq \left( \frac{R_{3}}{D}\right)^{\frac{10d_s}{3(3-d_s)}} \\
\simeq \left( \frac{a}{D} \right) ^{\frac{10d_s}{3(3-d_s)}} M^{\frac{2(2+d_s)}{3(3-d_s)}}
\label{statistically_branch_F_capillary}
\end{eqnarray}
For linear polymers ($d_s=1$), one again rocovers the well-known result of Eq. (\ref{F_slit}), {\it i.e.}, the free energy is proportional to the molecular weight.

{\it Confinement in a cavity}:
The free energy of branched polymers within a spherical cavity is obtained in a similar way.
Now, one has to replace Eq. (\ref{statistically_branch_blob_capillary}) by the following;
\begin{eqnarray}
\frac{g}{\xi^3} \simeq \frac{M}{D^3}
\label{statistically_branch_blob_sphere}
\end{eqnarray}
The blob size is given by
\begin{eqnarray}
\xi &\simeq& \left( \frac{D^{3(d_s+2)}}{R_{3}^{5 d_s}}\right)^{\frac{1}{2(3-d_s)}} \\
&\simeq& \left(\frac{D^{3(d_s+2)}}{a^{5d_s} M^{d_s+2}}\right)^{\frac{1}{2(3-d_s)}}
\label{sratistically_branch_xi_sphere_1}
\end{eqnarray}
By substituting $d_s = 1$, one recovers Eq. (\ref{correlation_length}) with $\nu_3 = 3/5$.
One sees that the correlation length now decays with the increase in $M$ even for linear polymers.
For the confinement free energy, one finds
\begin{eqnarray}
\frac{F_0}{k_BT} \simeq \frac{D^3}{\xi^3} &\simeq& \left( \frac{R_{3}}{D}\right)^{\frac{15d_s}{2(3-d_s)}} \\
&\simeq& \left( \frac{a}{D} \right) ^{\frac{15d_s}{2(3-d_s)}} M^{\frac{3(d_s+2)}{2(3-d_s)}}
\label{statistically_branch_F_sphere}
\end{eqnarray}
which is also deduced from the scaling argument (Eq. \ref{F_sphere}).
By substituting $d_s=1$, one recovers Eq. (\ref{F_sphere_good}).

The extension to branched polymers with weak branching density is discussed in the Appendix \ref{app1}.

\section{Injection of polymeric fractals into narrow capillary}
Up to now, our attention has been focused on equilibrium aspects of confined polymers.
One important fact is that the partition coefficient shows a crossover behaviour when the characteristic size of the confinement $D$ becomes comparable to the natural size of the coil $R_{3}$ in a bulk.
Although, as we already discussed, the sharpness of this crossover (dependence on the molecular weight) depends on the confinement regime, the common feature is that the probability to find the polymer in the space of $D < R_{3}$ is exponentially small.

However, it is possible to force the polymer into the space of size $D < R_{3}$ by applying certain flow, and the notion of SCR becomes much more evident here \cite{Sakaue_EPL}.
Aside from the fundamental interest, this is a subject of technological importance as a possible efficient methodology for the molecular characterization and separation \cite{deGennes_review}.
This section is devoted to this problem of the forced injection of polymeric fractals into narrow space.
The question of our interest is what is the minimum strength of solvent current $J$ (measured as solvent flux inside the capillary) to achieve the polymer injection.
One may naturally expect that the stronger current is necessary to inject larger branched polymers \cite{Gay,deGennes_review}.
We demonstrate, however, this is indeed wrong.

\begin{figure}[ht]
\begin{center}
\includegraphics[width=8cm]{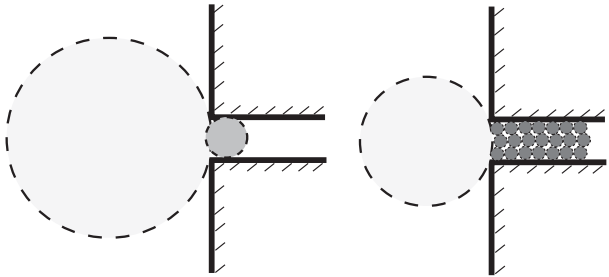}
\caption{Schematic drawing of the penetration process of a polymeric fractal into narrow capillary. The branched polymer with ($1 < d_s$) is progressively compressed with the penetration, consequently the blob size becomes smaller and smaller. The colour strength represents the density.}
\label{branch_tube}
\end{center}
\end{figure}

Confined in the narrow capillary, the branched polymer is in a SCR.
Therefore, the volume fraction of the branched polymer with ($1 < d_s$) increases progressively with the penetration process advanced as is schematically shown in Fig. \ref{branch_tube}.
(Such a feature is, in fact, of a central importance in the present problem, but missed in the past theory.)
Suppose that a fraction of the branched polymer is already sucked inside the capillary (Fig. \ref{branch_tube}).
We denote the length of the capillary section occupied by the polymer as $l$, and the number of monomers sucked inside as $P$.
To evaluate static structures of this polymer, one can follow the argument for the polymeric fractal in a capillary (Sec. \ref{confine_branch}), but now with the caution that our polymer is only partially injected.
Therefore, two quantities $l$ and $P$ are related by Eq. (\ref{statistically_branch_R_F_1_1}) with the replacement of $M$ by $P$.
\begin{eqnarray}
l \simeq \left[ \left( \frac{a^{5}}{D^{2}} \right) P^{\frac{d_s+2}{d_s}} \right]^{1/3}
\end{eqnarray}
Instead of Eq. (\ref{sratistically_branch_xi_capillary_1}), the blob size is now dependent on $l$
\begin{eqnarray}
\xi \simeq \left[ \frac{D^{2(d_s+2)}}{a^{5(d_s-1)}P^{\frac{(d_s-1)(d_s+2)}{d_s}}}\right]^{\frac{1}{3(3-d_s)}}
\end{eqnarray}
The number of monomer $g(l)$ inside a blob is related to $\xi(l)$ as the same relation for $g$ and $\xi$ (Eq. (\ref{sratistically_branch_xi_g_1})).
Whether this polymer will be further injected or pushed back is in view of the balance between the osmotic force and the hydrodynamic drag force.

The osmotic force due to the confinement is
\begin{eqnarray}
f_{osm} \simeq \Pi (l) D^2
\end{eqnarray}
where the osmotic pressure is given by $\Pi (l) \simeq k_BT/(\xi(l)^3)$.
The hydrodynamic force is evaluated to be a sum of Stokes drag force per blob.
\begin{eqnarray}
f_{hyd} \simeq \eta \xi(l) v \frac{P(l)}{g(l)}
\end{eqnarray}
where the viscosity of the solvent is denoted as $\eta$ and the velocity of the solvent inside a capillary is $v \simeq J/D^2$.
By balancing these forces, we see that the current should be larger than some threshold $J_c^{(l)}$ to further inject the polymer;
\begin{eqnarray}
\frac{\eta}{k_BT} J_c^{(l)}  \simeq \left[\left(\frac{D}{a}\right)^{5} P(l)^{-\frac{2+d_s}{d_s}}\right]^{\frac{2(2-d_s)}{3(3-d_s)}}
\label{J_c_l}
\end{eqnarray}
This is a decreasing function of $P$ (thus, $l$), which means that the more penetration process proceeds, the easier the pushing the branched polymer into a capillary is.
The crucial moment is the injection of a first blob of size $\xi(D)=D$ (Fig. \ref{branch_tube} left), therefore, we reach the critical current
\begin{eqnarray}
\frac{\eta}{k_BT} J_c = \frac{\eta}{k_BT} J_c^{(D)} \simeq 1
\label{J_c}
\end{eqnarray}
In contrast to our naive expectation that larger branched polymer would be injected into narrower capillary with more and more difficulties, the critical current does neither depend on molecular size $M$ nor the capillary size $D$, and noticeably it is the same as that for a linear polymer \cite{deGennes_review,Sakaue_EPL}.

The extension to the case of weaker branching (cf. Appendix \ref{app1}) is straightforward, and one can show that critical current does not depend on both molecular size $M$ and the degree of branching (in the notation of Appendix \ref{app1}, this is related to $b$), and again we are led to Eq. (\ref{J_c}).

\section{Conclusion}
In the present article, we have revisited the problem of the behavior of a confined polymer chain and introduced
the distinction between two confinement regimes.
In the first regime, called the {\it weak confinement regime} (WCR), the confinement leads to a reduction of the configurational entropy of the chain.
On the other hand, in the {\it strong confinement regime} (SCR), the chain becomes more and more compressed as its length
increases, and, as a result, the dominant contribution to the confinement free energy arises from the inter-segmental interactions.
This leads to some unique features of the SCR: (i) progressive penetration process, (ii) non-linear dependence of the confinement free energy on the segment number, (iii) local analogy with semidilute solutions, (iv) existence of a minimum allowable size of the confinement.
These features are closely related to each other, and the appearance of one of them is a
manifestation of the SCR.

For a linear chain, the critical geometry separating the above mentioned two regimes is 
the capillary tube. A linear chain confined in a closed cavity is thus in the SCR. We think that this may provide a possible interpretation of the recent experimental findings concerning 
the partitioning of PEG molecules into protein pores \cite{aHL,aHL2,aHL3}.

We have developed similar arguments for branched polymers with arbitrary connectivity.
We have shown in particular that branched polymers with spectral dimension $d_s >1$
are in a SCR already when confined into a capillary.
Using scaling arguments and a blob picture, we have analyzed some thermodynamic properties
(like the confinement free energy and correlation length) of polymeric fractals in various confinement geometries.

We have also discussed the problem of forced injection of polymeric fractals into a narrow capillary.
We have found that the critical current required for penetration depends neither on the molecular weight of the polymer nor on
 the capillary size. This result suggests that the simple forced permeation method discussed here will not be applicable  in the context of the technological application (such as molecular characterization and separation).
Nevertheless, we expect that by exploiting the nature of the SCR, it would be possible to construct an injection device which will enables us to probe 
the properties of branched structures.

Very recently, Nakaya {\it et al.} observed morphological changes in micro-emulsion droplets containing polymer chains
 \cite{Nakaya}. We expect that the notion of the strong confinement discussed in the present paper 
 plays a central role to interpret these experimental data.

\acknowledgments
We are grateful to F. Brochard-Wyart and P.-G. de Gennes for very fruitful discussions. This research was supported, in part, by JSPS Research Fellowships for Young Scientists (No. 4990). We have also benefited from interesting discussions with the members of the  ACI Nanosciences - Extrusion mol\'eculaire program (French Ministry of Education, Higher education and Research).
\appendix

\section{Confined branched object: case of weaker branching}
\label{app1}
In the main text, we assumed our branched polymer to be highly branched.
However, there also exist branched polymers whose branching density is lower, {\it i.e.}, there are substantial difunctional monomers between two adjacent branching point.
In this appendix, we extend our discussion on the confined polymers to the case of weaker branching using one of the most important examples of polymeric fractals, {\it i.e.}, {\it statistically branched polymer}.
In terms of spectral dimension, statistically branched polymers are characterized by $d_s \simeq 4/3$.
If one synthesizes branched polymers from the mixture of larger amount of difunctional monomers with multifunctional monomers, resultant products have lower branching density, then we denote the average number of such difunctional monomers between branching point as $b$.
The natural size of such an object in a bulk solution is \cite{Daoud_Joanny}
\begin{eqnarray}
R_{3} \simeq a M^{\frac{1}{2}}b^{\frac{1}{10}}
\label{statistically_branch_R_2}
\end{eqnarray}
The case with $b=1$ corresponds to the highly branched polymer.
Note that by substituting $b=M$, we recover the usual coil size of linear polymers in good solvent.

Some modification for the free energy of confined branched polymer is necessary accordingly.

{\it Chain in a slit}:
In the slit problem, the modification is straightforward.
From Eq. (\ref{statistically_branch_F_slit}), 
\begin{eqnarray}
\frac{F_2}{k_BT} \simeq \left(\frac{R_{3}}{D}\right)^2 \simeq \left( \frac{a b^{1/10}}{D}\right)^2 M
\end{eqnarray}

{\it Chain in a capillary}:
In this case, the modification is also straightforward if the diameter of the capillary is larger than some threshold $D^* = a M^{1/8}b^{19/40}$ \cite{Gay,deGennes_review}.
All we have to care is that now the blob size $\xi$ is related to the number of monomers inside $g$ as
\begin{eqnarray}
\xi \simeq a g^{\frac{1}{2}}b^{\frac{1}{10}}
\label{sratistically_branch_xi_g_2}
\end{eqnarray}
instead of Eq. (\ref{sratistically_branch_xi_g_1}).
Then, the correlation length is
\begin{eqnarray}
\xi \simeq \left( \frac{D^4}{R_{3}} \right)^{\frac{1}{3}} \simeq \frac{D^{4/3}}{a^{1/3}M^{1/6} b^{1/30}} \quad (D > D^*)
\label{sratistically_branch_xi_capillary_2}
\end{eqnarray}
and the confinement free energy is
\begin{eqnarray}
\frac{F_1}{k_BT} &\simeq& \frac{D^2 R_{1}}{\xi^3} \simeq \left( \frac{R_{3}}{D}\right)^{\frac{8}{3}} \\
&\simeq& \left( \frac{ab^{1/10}}{D} \right) ^{\frac{8}{3}} M^{\frac{4}{3}} \quad (D > D^*)
\end{eqnarray}
All these results coincides with those in the previous subsection if the distance between branching points is unity ($b=1$).
We call such a situation ($D>D^*$) as regime I\footnote{In ref. \cite{Gay}, regime I and regime II (see below) were called weak and strong confinement, respectively. However, it should be noticed that the meaning of these terminologies is different from the concept in our present discussion on the nature of polymer confinement.}.

However, if the capillary becomes narrower ($D<D^*$), additional considerations are necessary, since, then, the blob size $\xi$ (Eq. (\ref{sratistically_branch_xi_capillary_2})) becomes smaller than the characteristic spatial size of linear segment between branching points $\simeq a b^{3/5}$ \cite{Gay,deGennes_review}.
In such a case (called regime II), the confined branched object locally looks like semidilute solution of linear polymers, where, instead of Eq. (\ref{sratistically_branch_xi_g_2}), the relation $\xi \simeq a g^{3/5}$ is expected.
Therefore, the correlation length and the confinement free energy are, respectively, deduced as
\begin{eqnarray}
\xi \simeq D \left( \frac{b}{M}\right)^{\frac{1}{8}} \quad (D < D^*)
\end{eqnarray}
\begin{eqnarray}
\frac{F_1}{k_BT} &\simeq& \frac{D^2 R_{1}}{\xi^3} \simeq \left( \frac{R_{3}}{D}\right)^{\frac{5}{3}} \left(\frac{M}{b}\right)^{\frac{3}{8}}\\
&\simeq& \left( \frac{a}{D} \right) ^{\frac{5}{3}} \frac{M^{\frac{29}{24}}}{b^{\frac{5}{24}}} \quad (D < D^*)
\end{eqnarray}
By replacing $b$ by $M$ in these results in regime II, we recover results for a linear polymer in the capillary ($\xi = D$ and Eq. (\ref{F_slit})).

{\it Chain in a spherical cavity}:
The situation is the same as that in the capillary problem.
The correlation length in regime I and regime II are derived in a similar way as in the capillary problem, which are respectively written as 
\begin{eqnarray}
\xi \simeq \frac{D^3}{R_{3}^2} \simeq \frac{D^3}{a^2 M b^{\frac{1}{5}}} \quad (D > D^*)
\end{eqnarray}
\begin{eqnarray}
\xi \simeq \left( \frac{D}{a} \right)^{\frac{5}{4}} \frac{D}{M^{\frac{3}{4}}} \quad (D < D^*)
\end{eqnarray}
where the crossover cavity size is $D^* \simeq a M^{1/3}b^{4/15}$.
Therefore, the confinement free energy in these regimes are respectively
\begin{eqnarray}
\frac{F_0}{k_BT} &\simeq& \left( \frac{R_{3}}{D}\right)^{6} \\
&\simeq& \left( \frac{a}{D} \right) ^{6} b^{\frac{3}{5}}M^{3} \quad (D > D^*)
\end{eqnarray}
\begin{eqnarray}
\frac{F_0}{k_BT}  \simeq \left( \frac{a}{D} \right) ^{\frac{15}{4}} M^{\frac{9}{4}} \quad (D < D^*)
\end{eqnarray}
Both results of correlation length and free energy in regime II do not depend on $b$, and coincide with those for a linear polymer in a cavity (Eq. (\ref{correlation_length}) with $\nu_3 = 3/5$ and (\ref{F_sphere_good})).


\begin{thebibliography}{1}
\bibitem{deGennes}P.-G. de Gennes, {\it Scaling Concepts in Polymer Physics} (Cornell University Press, Ithaca, 1979).
\bibitem{Vilgis_2}T. A. Vilgis, Physics Reports, {\bf 336}, 167 (2000).
\bibitem{nanopore_exp3}{\it Structure and Dynamics of Confined Polymers}, edited by J.J. Kasianowicz {\it et al}. (Kluwer Academic Publishers, Dordrecht, 2002).
\bibitem{Kasiano}J. J. Kasianowicz, E. Brandin, D. Branton and D.ÊW. Deamer, Proc. Natl. Acad. Sci., {\bf 93},
13770 (1996).
\bibitem{LubNelson}D. K. Lubensky and D. R. Nelson, Biophys. J., {\bf 77}, 1824 (1999).
\bibitem{Mutu}M. Muthukumar, Phys. Rev. Lett., {\bf 86}, 3188 (2001)
\bibitem{MellerNivon}A. Meller, L. Nivon and D. Branton, Phys. Rev. Lett., {\bf 86}, 3425 (2001).
\bibitem{Chuang}J. Chuang, Y. Kantor and M. Kardar, Phys. Rev. E, {\bf 65}, 11802 (2002).
\bibitem{Alberts}B. Alberts, A. Johnson, J. Lewis, M. Raff, K. Roberts and P. Walter,
{\it{Molecular Biology of the Cell}}, Garland Publishing; 4th edition (March, 2002).
\bibitem{Daoud_deGennes}M. Daoud and P.-G. de Gennes, J. Phys. France, {\bf 38}, 85 (1977).
\bibitem{Cassasa}E. F. Cassasa, J. Polymer Sci. {\bf B5}, 773 (1967).
\bibitem{RedBook}A. Yu. Grosberg and A. R. Khokhlov, {\it  Statistical Physics of Macromolecules} (American Institute of Physics, New York, 1994).
\bibitem{aHL} S. M. Bezrukov, I. Vodyanoy, R. A. Brutyan and J. J. Kasianowicz, Macromolecules, {\bf 29}, 8517 (1996).
\bibitem{aHL2}P. G. Merzlyak, L. N. Yuldasheva, C. G. Rodriques, C. M. M. Carneiro, O. V. Krasilnikov and S. M. Bezrukov, Biophys. J. {\bf 77}, 3023 (1999).
\bibitem{aHL3}T. K. Rostovtseva, E. M. Nestorovich and S. M. Bezrukov, Biophys. J. {\bf 82}, 160 (2002).
\bibitem{aHL4}L. Movileanu and H. Bayley, Proc. Natl. Acad. Sci. USA. {\bf 98}, 10137 (2001).
\bibitem{aHL5}L. Movileanu, S. Cheley and H. Bayley, Biophys. J. {\bf 85}, 897 (2003).
\bibitem{nanopore_exp2}D.W. Deamer and D. Branton, Acc. Chem. Res. {\bf 35}, 817 (2002).
\bibitem{Vilgis_1}T. A. Vilgis, P. Haronska and M. Benhamou, J. Phys. II France, {\bf 4}, 2187 (1994).
\bibitem{Gay}C. Gay, P.-G. de Gennes, E. Rapha\"el and F. Brochard-Wyart, Macromolecules {\textbf 29}, 8379 (1996).
\bibitem{deGennes_review}P.-G. de Gennes, Adv. Polymer Sci., {\bf 138}, 92 (1999).
\bibitem{Lifshitz}I. M. Lifshitz, A. Yu. Grosberg and A. R. Khokhlov, Rev. Mod. Phys. {\bf 50}, 683 (1978).
\bibitem{Muth3}C. Y. Kong and M. Muthukumar, J. Chem. Phys. {\bf 120}, 3460 (2004).
\bibitem{DNA_partitioning_model_pore}D. Nykypanchuk, H. H. Strey and D. A. Hoagland, Macromolecules, {\bf 38}, 145 (2005).
\bibitem{Pincus}E. Raphael and Pincus, J. Phys. II France, {\bf 2}, 1341 (1992).
\bibitem{Alexander}S. Alexander and R. Orback, J. Physique Lett., {\bf 1982}, 43 (L625).
\bibitem{Isaacson}J. Isaacson and T. C. Lubensky, J. Phys. Lett., {\bf 1980}, 41 (L649).
\bibitem{Daoud_Joanny}M. Daoud and J.-F. Joanny, J. Phys., {\bf 1981},42 (359).
\bibitem{Sakaue_EPL}T. Sakaue, E. Rapha\"el, P.-G. de Gennes, F. Brochard-Wyart, submitted to EPL.
\bibitem {Nakaya} K. Nakaya, M. Imai, S. Komura, T. Kawakatsu and N. Urakami, Europhysics Lett.,  in press (2005).
\end{thebibliography}
\end{document}